\documentclass{optica-article}

\journal{opticajournal} 

\articletype{Research Article}

\usepackage{lineno}

\usepackage{makecell}
\usepackage{multirow}
\usepackage{float}
\usepackage{pifont}
\usepackage{amsmath}
\usepackage{graphicx}

\begin{document}

\title{ Measuring magnetic field coil constants based on atomic magnetometry and fluxgate magnetometry}

\author{Ni ZHAO,\authormark{1} Lulu ZHANG,\authormark{1} Yongbiao YANG,\authormark{1} Junye ZHAO,\authormark{1} Jun HE,\authormark{1,2} and Junmin WANG\authormark{1,2,*}}
 
\address{\authormark{1}State Key Laboratory of Quantum Optics and Quantum Optics Devices, and Institute of Opto-Electronics, 
Shanxi University, Tai Yuan 030006, Shanxi Province, China\\
\authormark2Collaborative Innovation Center of Extreme Optics, 
Shanxi University, Tai Yuan 030006, Shanxi Province, China\\

\email{\authormark{*}wwjjmm@sxu.edu.cn}} 


\begin{abstract*} 
In a magnetic field detection system, to achieve high-sensitivity magnetic field measurement, it is necessary to use uniform magnetic field coils to provide a stable working environment, so the measurement of the magnetic field coil’s constant is of great significance. To accurately measure the magnetic field and compare the coil’s constant, we employed two different methods under good magnetic shielding 
conditions: the optically-pumped rubidium free-induction decay (FID) magnetometry and the fluxgate magnetometry. In terms of measuring coil’s constant, the optically-pumped rubidium FID magnetometer performs better than fluxgate magnetometer due 
to its high-sensitivity and good signal-to-noise ratio (SNR). We compare the magnetic field measured by the FID magnetometer with that of the fluxgate magnetometer and obtain a calibration factor of 0.9967. The calibration of fluxgate magnetometer with optically-pumped atomic magnetometer is realized. The method is simple and easy to operate for calibrating fluxgate magnetometer.

\end{abstract*}

\section{Introduction}
A magnetic field coils system composed of Helmholtz coil pairs or saddle coils is often required to generate uniform magnetic fields regions in many fields, such as atomic magnetometers$^{1-3}$, atomic gyroscopes$^{4-5}$, neutron spin filters$^{6-7}$, biomedicine$^{8-9}$, and fundamental physics research.
The accurate measurement of the magnetic coil’s constant is extremely important. The value of the residual magnetic field in the magnetic shielding cylinder  can be accurately obtained by measuring the coil constant, and the performance of the magnetic field-shielded can be evaluated.  At the same time, the accurate measurement of the magnetic coil's constant can further compensate the ambient magnetic field noise.  

There are many methods for measuring the coil constants. Traditionally, a fluxgate magnetometer probe has been placed in the center of a magnetic shield cylinder to measure the magnetic field. However, the fluxgate magnetometer  has the problem of zero point drift and position error of the fluxgate -probe, which affect the accuracy of the magnetic field measurement. Therefore, this method has limited accuracy in measuring coil constant. The magnetic field can also be measured by the response of atoms to the magnetic field in space. Optically-pumped atomic magnetometers can extract magnetic field information based on the interaction between light
and atoms, which have been widely used for precision measurement of weak magnetic field because of its high sensitivity and wide dynamic magnetic field measurement range $^{10-12}$. 

In recent years, atomic magnetometers with high sensitivity have been used to calibrate magnetic coil constants. E. brechi $\emph{et al.}^{13}$ proposed an in-situ calibration method for a three-axis coil system based on the FID of spin-aligned atoms. H. Zhang  $\emph{et al.} ^{14}$used the Larmor precession frequency of a hyper polarized $^3$He atomic ensemble to calibrate the coil constants based on the spin-polarized co-magnetometer. L. Chen $\emph{et al.} ^{15}$ can calculate the duration of $\pi / 2$ pulse based on measuring the inert gas FID signal, and calibrate the coil constants. Q. Zhao $\emph{et al.} ^{16}$ through nondestructive phase measurement and coherent light pumped, the spin polarization of rubidium atoms is coherently regenerated, the Larmor precession signal is continuously oscillated, and the coil constants is calibrated.

In a previous study, we characterized the current noise of different commercial constant-current sources based on an optically -pumped FID atomic magnetometer$^{17}$.  The current noise level of the constant current source(Keysight model B2961A) is the lowest among the six tested constant current sources owing to the low-noise and high-stability.  To avoid excessive current noise affecting the magnetic field measurement accuracy, we choose the constant current source B2961A to output current to the  coils, and then the magnetic field will be generated in the center of the magnetic field shielding. The  accuracy of coil's constants are evaluated and compared by the optically-pumped rubidium FID magnetometer and the fluxgate magnetometer.

\section{Experimental Setup}
In our experiment, the magnetic field coils system and the atomic vapor cell are placed in a four-layer cylindrical high permeability permalloy magnetic field shield to suppress the environmental noise. In order to generate uniform magnetic field and reduce the interference between the coils and the shielding layers, the torque-free coils with low temperature coefficient and good rigidity are chosen to provide a magnetic field along the y direction. The constant current source B2961A is used to drive magnetic field coils along the y direction to generate magnetic field. 
A cubic-shaped atomic vapor cell (15 mm × 15 mm × 15 mm) containing $^{87}$Rb atoms and 100 Torr N$_2$ is used in our experiment. 

The experimental setup of the atomic magnetometer is shown in Figure 1.  The pump beam is emitted from a distributed Bragg reflector (DBR) laser, which is tuned to the $^{87}$Rb D1 transition line at 795 nm(from 5$^{2}$$\emph{S}_{1/2}$ $\emph{F}$=2 to 5$^{2}$$\emph{P}_{1/2}$ $\emph{F'}$=1) using saturated absorption spectroscopy (SAS) for frequency locking. The pump beam passes through two acousto-optic modulators (AOM1, AOM2). The AOM1 is mainly used to stabilize pump beam power. The AOM2 is used for switching control of the pump beam and compensate the frequency shift generated by the AOM1.
The on-time of the pump laser is 116 ns and the off-time is 134 ns. The pump beam is expanded by a telescope system (the beam diameter after expansion is about 10 mm), passes through a $\lambda / 4$  wave plate, becomes circularly polarized light, and enters the atomic gas chamber along the y direction. The linearly polarized probe laser, originated from a 780 nm DBR laser, is blue detuned by 6 GHz from the $^{87}$Rb D2 transition line at 780 nm(from 5$^{2}$$\emph{S}_{1/2}$ $\emph{F}$=1 to 5$^{2}$$\emph{P}_{3/2}$ $\emph{F'}$=2). The probe  beam has a diameter of 2 mm and a power of 30 $\mu$W. The direction of the probe beam is perpendicular to the pump beam and the RF magnetic field. The probe beam passes through the atomic vapor cell and enters the polarimeter composed of a $\lambda / 2$  wave plate, a Wollaston prism, and a balanced differential photoelectric detector. We obtain information about the Faraday rotation angle by the photoelectric detector. The NI Data Acquisition (DAQ) card (NI-USB6363) and LabVIEW comprise the data acquisition system. The pump laser, the RF magnetic field and the probe laser are separated in the time domain by time sequence control to avoid the crosstalk effect on the measurement signal and sensitivity, and further influence on the coil's constants characterization.

\begin{figure}[h]
    \centering
    \includegraphics[width=0.9 \textwidth,height=0.55\textwidth]{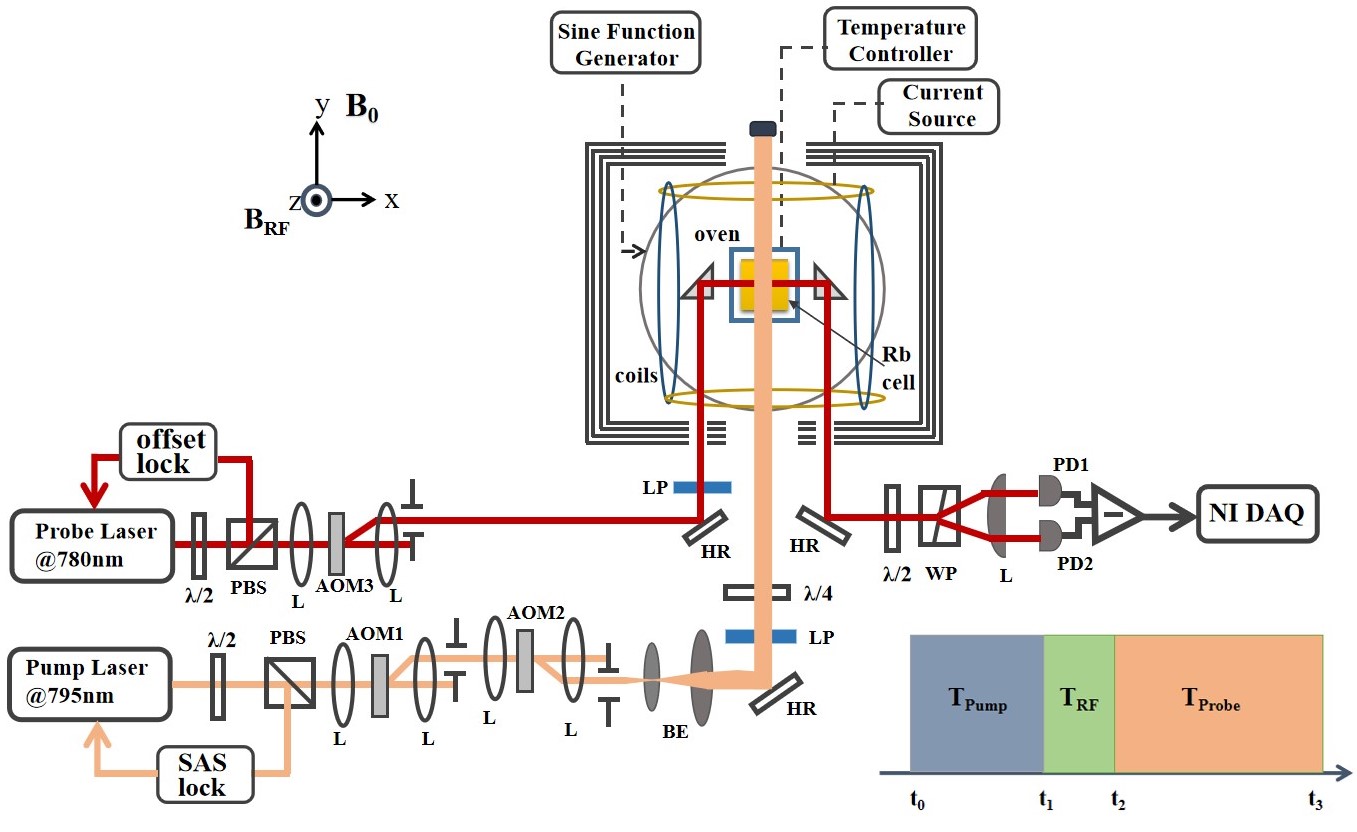}
    \caption{ Experimental setup of a  optically-pumped rubidium free induction decay(FID) magnetometer. SAS lock: saturated absorption spectroscopy for the pump laser frequency lock; Offest lock: the probe laser frequency stabilization based on a confocal Fabry-Perot cavity; AOM: acousto-optical modulator; BE: beam expander; $\lambda / 4$ : quarter-wave plate;  $\lambda / 2$ : half-wave plate; LP: linear polarizer with a high extinction ratio;  WP: Wollaston prism;
     L:lens; PD1(PD2): balanced differential photoelectric detector; 
     NI DAQ: data acquisition.}
   
\end{figure}

\section{Measurement results and discussion}
\subsection{ Magnetic field measurement and error}

Figure 2 shows the results of measuring magnetic fields based on optically-pumped FID rubidium magnetometer. The period T of FID signalis 50 ms. The pump beam is switched for 20 ms. The RF magnetic field is on for 0.2 ms. The probe beam is switched for 29.8 ms. We apply a current of 100 mA along the y direction. The heating temperature of the atomic vapor cell is set at 85°C, and the atomic number density is about 2.2 × 10$^{12}$ cm$^3$. Figure 2(a) shows a typical FID signal in one period. Figure 2(b) shows the Fast Fourier Transform (FFT) signal of FID signal. A full width at half maximum(FWHM) is about 337.2±1.9 Hz. The central frequency is the Larmor precession frequency, and the magnetic field can be calculated by the gyromagnetic ratio of ground state $^{87}$Rb atoms. Figure 2(c) shows the 4800 DC magnetic field values obtained through calculation. According to the statistical average of the magnetic field values distribution, the static magnetic field is approximately 12.62859 µT.  Figure 2(d) shows the power spectral density (PSD) calculated with the magnetic field values, which shows a magnetic sensitivity of 6.7 pT/Hz$^{1/2}$ with a bandwidth of 0.5–10 Hz. The peak is caused by the 50-Hz electronic noise and its harmonic.

\begin{figure}[h]
    \centering
    \includegraphics[width=0.9 \textwidth,height=0.7\textwidth]{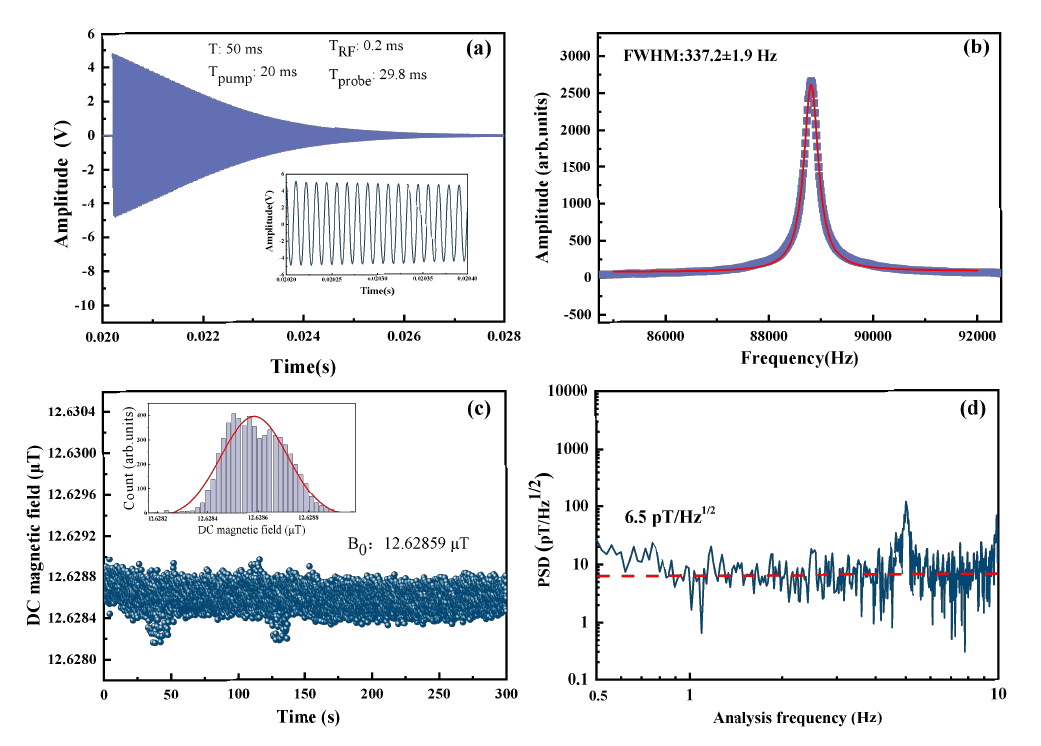}
    \caption{ (a) A typical FID signal for a period;(b) The FFT of the FID signal; (c) DC magnetic field values for 4800 sampling periods. (d) The PSD of magnetic field noise, which is about 6.7 pT/Hz$^{1/2}$ with a bandwidth of 0.5–10 Hz. }
   
\end{figure}

As shown in Figure 3, the fluxgate magnetometer(Model FVM400) used in our experiment is a three-axis magnetometer manufactured by Macintyre Electronic Design Associates( MEDA ) with a magnetic induction probe(the probe size is 25.4 mm×25.4 mm×100.6 mm), which measure magnetic fields in different direction, the measurement range of the magnetic fields is ±100000 nT. The permits resolution of 1 nT in a 100,000 nT field. In the experiment, we placed the of the probe the fluxgate magnetometer at the center of the coil bobbin
 in a magnetic shielding cylinder with good shielding performance for magnetic field measurement. We adjust the position and angle of the probe of the magnetometer so that the y axis of the probe of the fluxgate magnetometer is completely parallel to the y axis of the magnetic field coil. We used the fluxgate magnetometer to measure the magnetic fields along the y axis. Before each measurement of the magnetic field, the
 fluxgate magnetometer needs to be zeroed, that is, the magnetic probe is zeroed in a magnetic shielding cylinder. This magnetic shield provides good zero-field conditions with a residual magnetic field of only a few Nano Tesla. The ultra-low noise and high precision constant current source provide a very stable static magnetic field to ensure the reliability of the experiment.

 \begin{figure}[h]
    \centering
    \includegraphics[width=0.4\textwidth,height=0.5\textwidth]{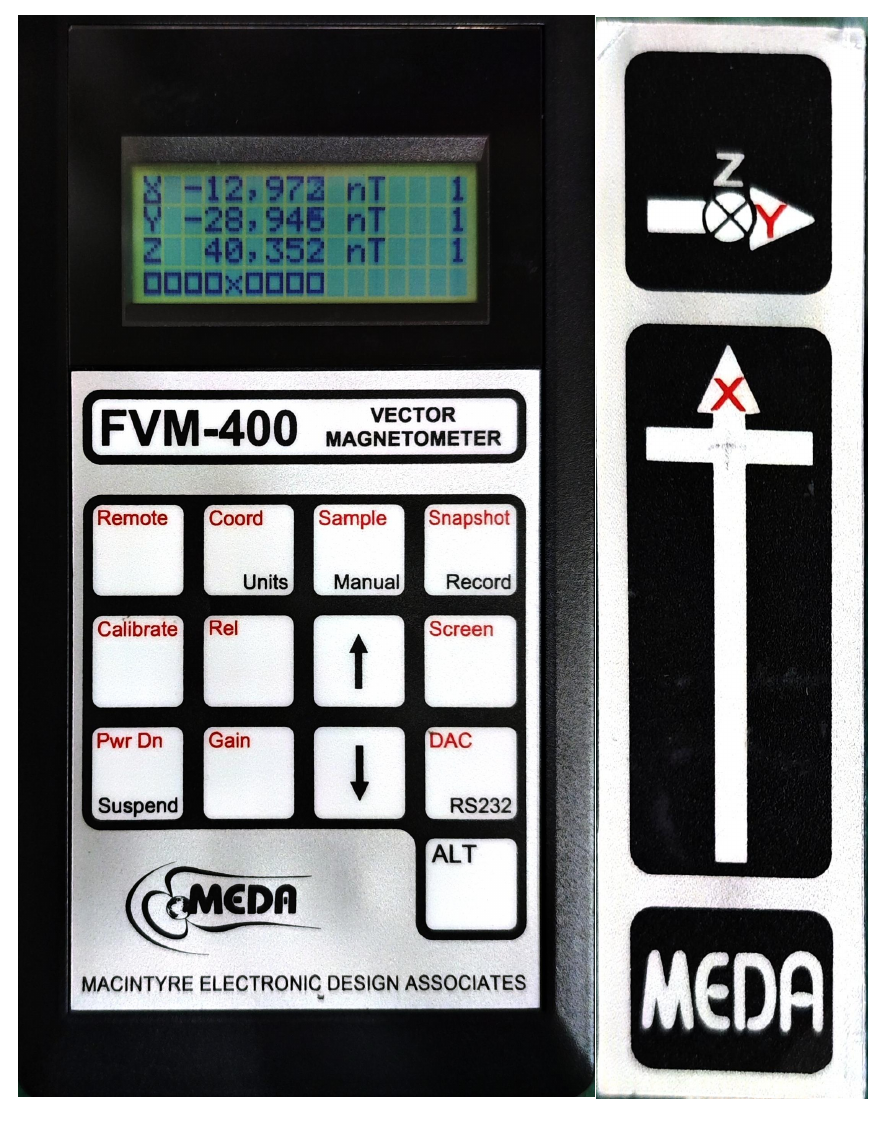}
    \caption{ Model FVM400 three-axis  fluxgate magnetometer  (Macintyre Electronic Design Associates, MEDA, USA).}
   
\end{figure}
 
The probe of the fluxgate magnetometer is placed in the original position of the atomic vapor cell, and the measurement results of the fluxgate magnetometer are evaluated and calibrated using the optically-pumped FID rubidium magnetometer. The same currents were applied to the magnetic field coils, and measurement results of the fluxgate magnetometer are compared with the magnetic field obtained by using optically-pumped FID rubidium magnetometer. Table 1 compares the magnetic field
 measurement results and errors based on the two methods under the same current applied by a constant current source, which show that the measurement results of the fluxgate magnetometer are slightly larger than the actual magnetic field value. It shows that the accuracy of the fluxgate magnetometer is not very high and its measuring deviation is large when measuring weak magnetic fields.

\begin{table*}
  \begin{center}
    \renewcommand{\arraystretch}{1.5}
 \caption{Comparison of magnetic fields measured by two methods under the same current applied by a constant current source.}
    \label{table_example}
    \centering
    \begin{tabular}{|c|c|c|c|c|c|}
       
       \hline
       Current(mA)  & 20& 50  & 100 &200 & 250\\
          \hline
   {\makecell{Magnetic field given \\by the Fluxgate\\magnetometer(nT)}} & {\makecell{2569\\
±3}}  & {\makecell{6367 \\±8}} &  {\makecell{12747\\±12}}   &  {\makecell{25501\\±13}}& {\makecell{31881\\±12} } 
        \\
      
         \hline
   {\makecell{Larmor frequency given \\by the FID atomic\\magnetometer(Hz)}} & {\makecell{17748.823\\
±0.632}}  & {\makecell{44419.960 \\±0.914}} &  {\makecell{88754.509\\±0.622}}   &  {\makecell{177562.703\\±0.438}}& {\makecell{221984.548\\±0.425} } 
        \\
       
         \hline
   {\makecell{Linewidth given by \\ the FID atomic\\magnetometer(Hz)}} & {\makecell{277.243\\
±2.312}}  & {\makecell{292.438 \\±2.914}} &  {\makecell{337.276\\±1.906}}   &  {\makecell{394.026\\±1.232}}& {\makecell{402.823\\±1.212} } 
        \\
        
         \hline
   {\makecell{Magnetic field given \\by the FID atomic\\magnetometer(nT)}} & {\makecell{2537.057\\
±0.090}}  & {\makecell{6349.491 \\±0.131}} &  {\makecell{12686.773\\±0.093}}   &  {\makecell{25381.220\\±0.063}}& {\makecell{31730.981\\±0.061} } 
        \\
         \hline
    \end{tabular}       
  \end{center}
\end{table*}

\subsection{Measurement of coil’s constant by using of an optically-pumped FID rubidium magnetometry and fluxgate magnetometry}

Figure 4(a) shows the experimental results of calibrating the coil constants based on the optically pumped FID atomic magnetometer. Figure 4(b) shows measurement error of magnetic field of FID atomic magnetometer at different current. At first, the constant current source apply a known current to the torque-free coils. Then we record the FID signal for 240 s. The Larmor frequency is obtained by FFT transformation, and magnetic field value is obtained by calculation and statistical averaging. The constant current source applies current in the range of 2-250 mA. A series of magnetic field values are measured by using a FID magnetometer. The linear fitting results can be obtained as follows:

\begin{flalign}
  \ B(nT)=126.956(nT/mA)I(mA)-4.914(nT)       
\end{flalign}

\begin{figure}[h]
    \centering
    \includegraphics[width=1.0 \textwidth,height=0.4\textwidth]{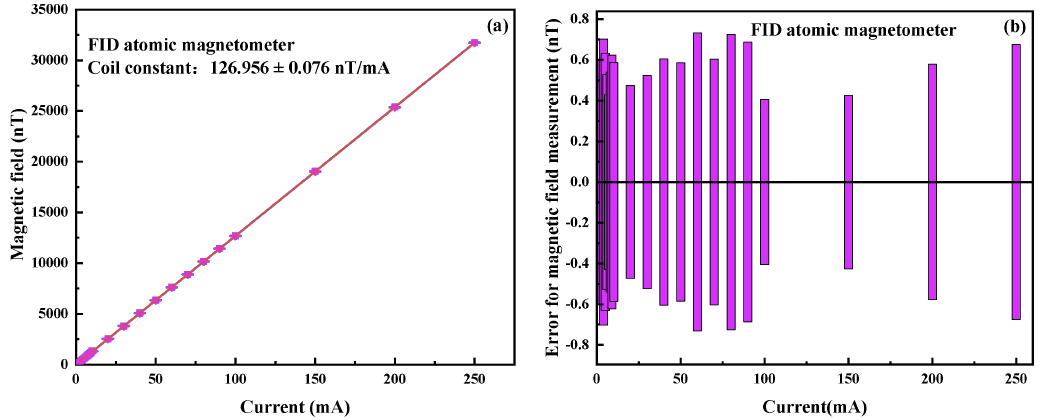}
    \caption{ (a) Result of the coil constant calibrated by using of FID atomic magnetometer. The coil constant is 126.956 ± 0.076 nT/mA. (b)Measurement error of magnetic field of FID atomic magnetometer at different current.}
  
\end{figure}
According to the fitted linear equation, the calibrated coil constant is approximately 126.956 ± 0.076 nT/mA. The residual magnetic field at the Rb cell position inside the four-layer magnetic fielding tank is about 4.914 nT.

Applying current within the range of 2-250 mA, the magnetic field values measured based on the fluxgate magnetometer are shown in Figure 5(a). Figure 5(b) shows measurement error of magnetic field of the fluxgate magnetometer at different current.
The fluxgate magnetometer record 525 DC magnetic field values for 7.5 seconds at the same current. Through the statistical average of magnetic field values, we can get the corresponding magnetic field values. After linear fitting, the relationship between the magnetic field  and current is obtained as:
\begin{flalign}
  \ B(nT)=127.3(nT/mA)I(mA)-4.914(nT)        
\end{flalign} 

As we know from the fitted equation, the value of the coil constant calibrated based on the fluxgate magnetometer is about 127.3±0.3 nT/mA.
\begin{figure}[h]
    \centering
    \includegraphics[width=1.0\textwidth,height=0.4\textwidth]{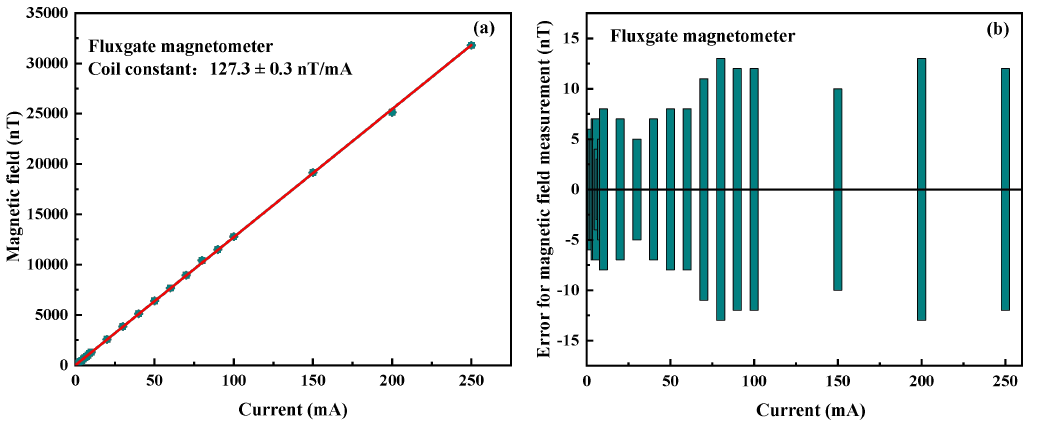}
    \caption{ (a) Result of the coil constant calibration by using a fluxgate magnetometer. The coil constant is 127.3±0.3 nT/mA. (b) Measurement error of magnetic field of fluxgate magnetometer at different current.}
   
\end{figure}

Table 2 compares the coil constants measured based on the two methods. The  experimental results show that compared with fluxgate magnetometer, the optically-pumped rubidium atomic magnetometer has the highest accuracy and smallest error in measuring the coil constant. The optically-pumped rubidium FID magnetometer has high-SNR, narrow-linewidth and high-sensitivity due to the preparation of spin polarization states by the pump laser for the atomic ensemble. The linewidth is in the order of hundreds of Hz. And the measurement process involves measuring multiple cycles and taking statistical averages, resulting in high accuracy and small error in the Larmor frequency. The fluxgate magnetometer is susceptible to external environment
and temperature, and has zero drift problems, which limits the accuracy of magnetic measurement, resulting in the worst calibration accuracy of coil constant.

\begin{table*}
    \renewcommand{\arraystretch}{1.5}
 \caption{Comparison of the two methods for calibrating magnetic field coil constants.}
    \label{table_example}
    \centering
    \begin{tabular}{|c|c|c|c|}
      
        \hline
         &  {\makecell{Using atomic \\magnetometer}} &  {\makecell{Using the fluxgate\\  magnetometer} }  \\
       \hline
      {\makecell{Coil constants\\( nT/mA )}}  &    126.956 ± 0.076 &   127.3 ± 0.3 \\  
        \hline
    \end{tabular}
\end{table*}

\subsection{Calibration of commercial fluxgate magnetometer based on optically-pumped FID atomic magnetometer}
In the experiment, we optimize the parameters, such as the temperature of rubidium vapor cell, the power of pump beam, and the strength of RF magnetic field to improve SNR of magnetic resonance signal and reduce its linewidth. We measured the magnetic field separately based on fluxgate magnetometers and optically-pumped FID atomic magnetometers under the same current. After fitting, the current applied by the constant current source is linearly related to the magnetic field generated by the torque-free coils pairs. The coil constant k$_{1}$ is 126.956±0.076 nT/mA obtained by the FID atomic magnetometer. The coil constant k$_{2}$ is 127.3±0.3 nT/mA obtained by the fluxgate magnetometer. We defined calibration factor C to calibrate the measurement results of the fluxgate magnetometer, where C=k$_{2}$/k$_{1}$.  As shown in Figure 6, we obtained a calibration factor of 0.96967. We define B’=B$_{0}$*C, where B$_{0}$ is the magnetic field value of the fluxgate magnetometer. Therefore, the magnetic field value of a fluxgate magnetometer can be calibrated by an optically-pumped atomic magnetometer.

 \begin{figure}[h]
    \centering
    \includegraphics[width=0.6\textwidth,height=0.5\textwidth]{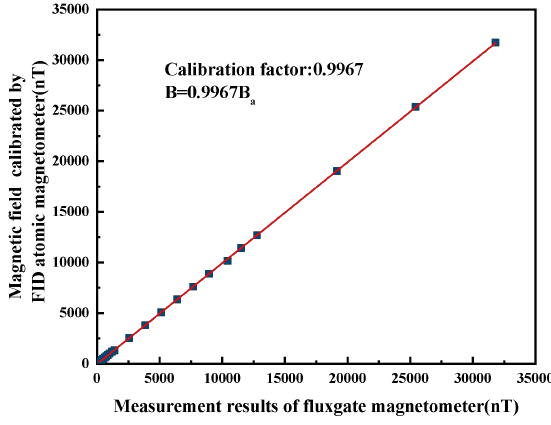}
    \caption{ The relationship between the magnetic fields calibrated by optically-pumped rubidium FID magnetometer and the measurement results of the fluxgate magnetometer.}
  
\end{figure}

\section{Conclusion}

    In this paper, we measure magnetic field and compare coil's constant based on the optically-pumped rubidium FID magnetometer and the fluxgate magnetometer. In terms of coil's constants calibration, the optically-pumped FID magnetometer is superior to the fluxgate magnetometer because of its high sensitivity and good SNR.  In addition, We demonstrated employing optically-pumped rubidium FID magnetometer to calibrate the fluxgate magnetometer when measuring magnetic field, which has been accomplished in a good magnetic shielding environment. The magnetic resonance signal with high-SNR and narrow linewidth is obtained by optimizing the parameters, and the typical magnetic resonance linewidth is hundred of Hz in the experiment. The magnetic field value can be accurately obtained based on the precisely positioned Larmor frequency and the gyromagnetic ratio of ground state atoms. As a result, the fluxgate magnetometer is calibrated by the optically-pumped rubidium FID magnetometer. 

   In the future, We can introduce squeezed-light to suppress the photon shot noise limit$^{19-20}$, further improve the sensitivity of the atomic magnetometer, which can make the measurement of the coil constant more accurate. In addition,
    based on the calibrated coil constant, we can carry out active magnetic field stabilization$^{21}$, so as to further compensate and shield the ambient magnetic field noise, and realize the development of high-sensitivity atomic magnetometers. Further, we can convert the magnetic field sensitivity to the current noise of constant current source by coil constant$^{22-24}$, achieve the characterization of the current noise of constant current source.

\section*{Data availability statement}
Data underlying the results presented in this paper are
not publicly available at this time but may be obtained from the authors upon reasonable request.

\section*{Acknowledgment} 
 National Key R\&D Program of China (2021YFA1402002); National Natural Science Foundation of China (11974226);  Shanxi Provincial Graduate Education Innovation Project (2022Y022).

\section*{Disclosures}
The authors declare no conflicts of interest.

\section* {ORCID iDs}
Junmin Wang https://orcid.org/0000-0001-8055-000X
\section*{References}
1.D. Budker and M. Romalis, "Optical magnetometry," Nat. Phys. {\bf3}(4), 227–234
(2007).\\
2.R. Zhang, D. Kanta, A. Wickenbrock, H. Guo, and D. Budker, "Heading-error free optical atomic magnetometry in the earth-field range," Phys. Rev. Lett. {\bf 130}(15),
153601 (2023).\\
3.S. R. Li, D. Y. Ma, K. Wang, Y. N. Gao, B. Z. Xing, X. J. Fang, B. C. Han, and W. Quan, "High sensitivity closed-loop Rb optically pumped magnetometer for
measuring nuclear magnetization," Opt. Express {\bf 30}(24), 43925–43937 (2022).\\
4.D. Meyer and M. Larsen, “Nuclear magnetic resonance gyro for inertial navigation,” Gyroscopy and Navigation. {\bf 5}(2), 75–82 (2014).\\
5. T. G. Walker and M. S. Larsen, "Spin-exchange-pumped NMR gyros," Adv. At. Mol. Opt. Phy. {\bf 65}, 373–401 (2016).\\
6.C. Y. Jiang, X. Tong, D. R. Brown, S. Chi, A. D. Christianson, B. J. Kadron, J. L. Robertson, and B. L. Winn, "Development of a compact in situ polarized $^3$He neutron spin filter at Oak Ridge National Laboratory," Rev. Sci. Instrum. {\bf 85}(7), 075112 (2014).\\
7.E. Karimi, L. Marrucci, V. Grillo, and E. Santamato, "Spin-to-orbital angular momentum conversion and spin-polarization filtering in electron beams," Phys. Rev. Lett. {\bf 108}(4), 044801 (2012).\\
8.M. E. Limes, E. L. Foley, T. W. Kornack, S. Caliga, S. McBride, A. Braun, W. Lee, V. G. Lucivero, and M. V. Romalis, "Portable magnetometry for detection of biomagnetism in ambient environments," Phys. Rev. Appl. {\bf 14}(1), 011002 (2020).\\
9.T. H. Sander, J. Preusser, R. Mhaskar, J. Kitching, L. Trahms, and S. Knappe,
"Magnetoencephalography with a chip-scale atomic magnetometer," Biomed. Opt. Express {\bf 3}(5), 981–990 (2012).\\
10.D. Sheng, S. Li,  S. Dural, and M. V. Romalis, "Subfemtotesla scalar atomic magnetometry using multipass cells," Phys. Rev. Lett. {\bf 110}(16), 160802( 2013).\\
11. D. Hunter,  S. Piccolomo, J. D. Pritchard ,  N. L.Brockie, T. E. Dyer, and  E. Riis, "Free-induction-decay magnetometer based on a microfabricated Cs vapor cell," Phys. Rev. A  {\bf 10}(1), 014002(2018). \\
12.C. Liu, H. F. Dong, and  J. J. Sang, "Submillimeter resolution magnetic field imaging with digital micromirror device and atomic vapor cell," Appl. Phys. Lett. {\bf 119}(11), 114002(2021). \\
13.E. Breschi,  Z. Grujic, and  A. Weis, "In situ calibration of magnetic field coils using free-induction decay of atomic alignment," Appl. Phys. B  {\bf 115,} 85-91(2014). \\
14.H. Zhang, S. Zou, and  X. Y. Chen, "A method for calibrating coil constants by using an atomic spin co-magnetometer," European Phys. J. D  {\bf 70,} 1-5(2016). \\
15.L. L. Chen,  B. Q. Zhou, G. Q. Lei,  W. F. Wu, J. Wang, Y. Y. Zhai,  Z. Wang, and  J. C. Fang, "A method for calibrating coil constants by using the free induction decay of noble gases," AIP Advances {\bf 7}(7), 075315(2017). \\
16.Q. Zhao,  B.  L. Fan, S. G. Wang, and L. J. Wang,  "A calibration method for coil constants using an atomic spin self-sustaining vector magnetometer," J. Magnetism and Magnetic Materials {\bf 514,} 166977(2020) . \\
17. N. Zhao, L. L. Zhang, Y. B. Yang, J. He, Y. H.  Wang, T. Y. Li, and J. M. Wang, "Characterizing current noise of commercial
constant-current sources by using an optically pumped rubidium atomic magnetometer"  Rev. Sci. Instrum. {\bf 94} 095001(2023).\\
18.L. L. Zhang, L. L. Bai , Y. L. Yang, Y. B. Yang,  Y. H. Wang, X. Wen, J. He, and J. M. Wang, "Improving the sensitivity of an optically pumped rubidium atomic magnetometer by using of a repumping laser beam," Acta Physica Sinica {\bf 70}(23), 230702(2021).(in Chinese) \\
19.C. Troullinou, R. Jiménez-Martínez, J. Kong,  V. G. Lucivero, and M. W. Mitchell, “Squeezed-light enhancement and backaction evasion in a high sensitivity optically pumped magnetometer,” Phys. Rev. Lett. {\bf 127}, 193601(2021).\\
20. L. L. Bai, X. Wen, Y. L. Yang, L. L.Zhang, J. He,  Y. H. Wang, and J. M. Wang, “Quantum-enhanced rubidium atomic magnetometer
based Faraday rotation via 795  Stokes operator squeezed light,” J. Opt. {\bf 23}, 085202(2021).\\
21.Y. D. Ding, R. Zhang, J. H. Zheng, J. B.  Chen, X. Peng, T. Wu, and H. Guo, “Active stabilization of terrestrial magnetic field with potassium atomic magnetometer,” Rev. Sci. Instrum. {\bf 93}(1), 015003(2022). \\
22.D. Y. Chen, P. X. Miao, Y. C. Shi, J. Z. Cui, Z. D. Liu, J. Chen, and K. Wang, “Measurement of noise of current source by pump-probe atomic magnetometer,” Acta Physica Sinica.  {\bf 71}(2), 024202 (2022). (in Chinese)\\
23.L. Shen, R. Zhang, T. Wu, X. Peng, S. Yu, J. B. Chen, and H. Guo, “Suppression of current source noise with an atomic magnetometer,” Rev. Sci. Instrum. {\bf 91}(8), 084701 (2020).\\
24.J. T. Zheng, Y. Zhang, Z. Y. Yu, Z. Q. Xiong, H. Luo, and Z. G. Wang, “Precision measurement and suppression of low-frequency noise in a current source with double-resonance alignment magnetometers,” Chinese. Phys. B.{\bf 32}(4), 040601(2023).\\

\end{document}